\documentclass[onecolumn,10pt]{article}

\usepackage{hyperref}   
\hypersetup{
	colorlinks=true,
	linkcolor=blue,
	citecolor=blue,
	urlcolor=blue,
}
\usepackage{amsmath}
\usepackage{amssymb}
\usepackage{tikz}
\usetikzlibrary{shapes,arrows,backgrounds,shadows,petri,calc}

\title{Sensitivity analysis using the\\ Metamodel of Optimal Prognosis}
\author{Thomas Most, Johannes Will, Dynardo GmbH, Weimar, Germany}
\date{\small 8th Optimization and Stochastic Days, Weimar, Germany, 24-25 November, 2011}

\begin{document}
\maketitle    

\section{Introduction}
Optimization and robustness analysis have become important tools for the virtual development of industrial 
products. In parametric optimization, the optimization variables are systematically 
modified by mathematical algorithms in order to get an improvement of an existing design or 
to find a global optimum. The design variables are defined by their lower and upper bounds
or by several possible discrete values. In real world industrial optimization problems, the number of design variables
can often be very large. Unfortunately, the efficiency of mathematical optimization algorithms decreases with increasing
number of design variables. For this reason, several methods are limited to a moderate number of variables,
such as gradient based and Adaptive Response Surface Methods.
With the help of sensitivity analysis the designer identifies the variables which 
contribute most to a possible improvement
of the optimization goal. Based on this identification, the number of design variables may be dramatically reduced and 
an efficient optimization can be performed. Additional to the information regarding important variables,
sensitivity analysis may help to decide, if the optimization problem is formulated appropriately and if the
numerical CAE solver behaves as expected.

By definition, sensitivity analysis is the study of how the uncertainty
in the output of a model can be apportioned, qualitatively or quantitatively, 
to different sources of variation in the input of a model \cite{Saltelli2000}.
Since robustness analysis investigates the influence of the input variation on the 
variation of the model outputs, sensitivity analysis can directly be applied as a post-processing tool
to analyze the contribution of each input variable to the scatter of each model response.
In order to quantify this contribution, variance based methods are very suitable.
With these methods, discussed in this paper, the proportion of the output variance, which is caused by an
random input variable, is directly quantified. 

For optiSLang robustness analysis, the scattering input variables are defined as random variables.
This means that for each scattering input a distribution type including mean value and variance is specified.
Additionally, dependencies between the inputs can be formulated in terms of linear correlations.
The model output variation is estimated by random sampling.
The estimated variation and the sensitivity measures are strongly influenced by the 
chosen variation of the input variables. 

Variance based sensitivity analysis is also very suitable as an optimization pre-processing tool.
By representing continuous optimization variables by uniform distributions without variable interactions,
variance based sensitivity analysis quantifies the contribution of the optimization variables to 
a possible improvement of the model responses.
In contrast to local derivative based sensitivity methods, the variance based approach quantifies
the contribution with respect to the defined variable ranges.
 
Unfortunately, sufficiently accurate variance based methods require huge numerical effort due to the large number of simulation runs.
Therefore, often meta-models are used to represent the model responses surrogate functions in terms of the model inputs.
However, many meta-model approaches exist and it is often not clear which one is most suitable for which problem \cite{Roos_2007_WOST}.
Another disadvantage of meta-modeling is its limitation to a small number of input variables. Due to the
curse of dimensionality the approximation quality decreases for all meta-model types 
dramatically with increasing dimension. As a result, an enormous number of samples is necessary to represent high-dimensional problems with
sufficient accuracy. In order to overcome these problems, Dynardo developed the Metamodel of Optimal Prognosis \cite{Most_2008_WOST}.
In this approach the optimal input variable subspace together with the optimal meta-model approach
are determined with help of an objective and model independent quality measure, the Coefficient of Prognosis.
In the following paper the necessity of such a procedure is explained by discussing other existing methods for sensitivity analysis.
After presenting the MOP concept in detail, the strength of this approach is clarified by a comparison with very common meta-model approaches
such as Kriging and neural networks.
Finally, an industrial application is given, where the benefit of the MOP is illustrated.

\section{Scanning the space of input variables}
In order to perform a global sensitivity analysis, the space of the input variables, which is either the design or the random
space, has to be scanned by discrete realizations. Each realization is one set of values belonging to the specified inputs.
For each set of values the CAE model is a black box solver and the model responses are evaluated.
In the case of random variables, only random sampling can be used to generate discrete realizations. 
The sampling scheme needs to represent the specified variable distributions and their dependencies with a sufficient accuracy.
A very common approach is Monte Carlo Simulation (MCS).
However, if only a small number of samples is used,
often clusters and holes can be remarked in the MCS sampling set. More critical is
the appearance of undesired correlations between the input variables. These correlations may have a significant influence on the
estimated sensitivity measures.
In order to overcome such problems, optiSLang provides optimized Latin Hypercube Sampling (LHS),
where the input distributions and the specified input correlations are represented very accurately even for a small number of
samples. For the minimization of the undesired correlation the method according to \cite{Iman1982} is used.

\begin{figure}[th]
\centering
\includegraphics[width=0.45\textwidth]{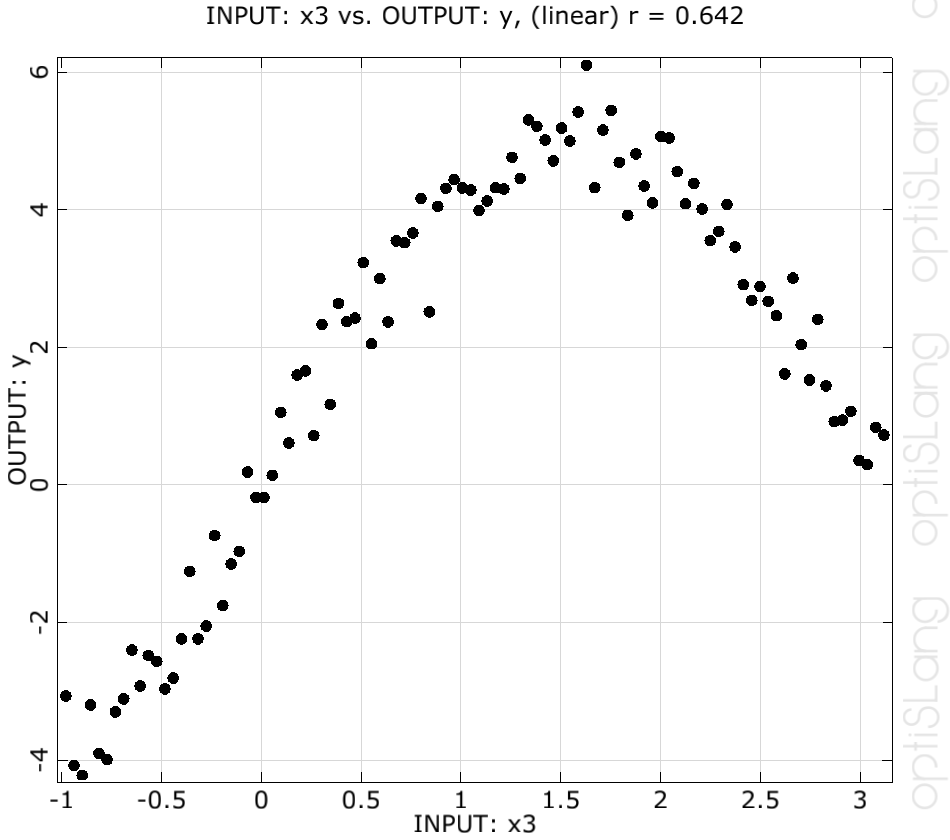}
\hspace{1cm}
\includegraphics[width=0.45\textwidth]{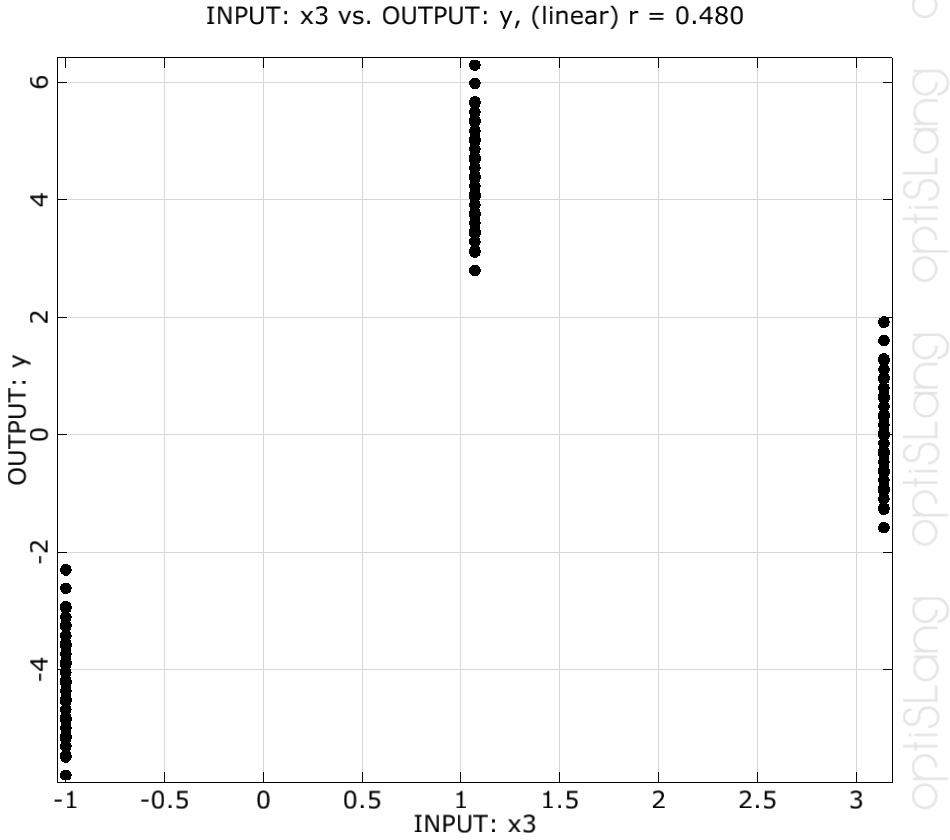}
\caption{Dimension reduction for a nonlinear function with five inputs 
based on Latin Hypercube Sampling (left) with 100 samples and full factorial design (right) with $3^5=243$ samples}
\label{DOE}
\end{figure}
As a design exploration for optimization problems deterministic Designs of Experiments (DoE) are often applied \cite{Myers2002}.
These design schemes are mainly based on a regular arrangement of the samples, as in the full factorial design. 
Generally the number of samples increases exponentially with increasing dimension.
Fractional factorial designs use only a part of the full factorial samples, however the number of levels in each direction is limited to three.
From our point of view, deterministic design schemes have two main disadvantages compared to random sampling:
They are limited to a small number of variables due to the rapidly increasing number of required samples when  increasing the model dimension.
Further a reduction of the number of inputs does not improve the information gained from the samples, since only two or three levels are used 
in each dimension. This is illustrated in Figure \ref{DOE}.
In this figure, a nonlinear function having one major and four almost unimportant variables is evaluated.
Using the LHS, the nonlinearity can be represented very well in the reduced space. In the case of the full factorial design, which contains three 
 levels in each directions, again only three positions are left in the reduced space and the dimension reduction does not allow a better representation of the 
model response.

\section{Variance based sensitivity analysis}
\subsection{First order and total effect sensitivity indices}
\label{saltelli}
Assuming a model with a scalar output $Y$ as a function of a given set of $m$ random input parameters $X_i$
\begin{equation}
Y=f(X_1,X_2,\ldots,X_m),
\label{model}
\end{equation}
the first order sensitivity measure was introduced as \cite{Sobol1993}
\begin{equation}
S_i=\frac{V_{X_i}(E_{\mathbf{X}_{\sim i}}(Y|X_i))}{V(Y)},
\label{first}
\end{equation}
where $V(Y)$ is the unconditional variance of the model output and 
$V_{X_i}(E_{\mathbf{X}_{\sim i}}(Y|X_i))$ is named the {\it variance of  conditional expectation}
with $\mathbf{X}_{\sim i}$ denoting the matrix of all factors but $X_i$.
$V_{X_i}(E_{\mathbf{X}_{\sim i}}(Y|X_i))$ measures the first order effect of $X_i$ on the model output.

Since first order sensitivity indices measure only the decoupled influence of each variable
an extension for higher order coupling terms is necessary.
Therefore total effect sensitivity indices have been introduced \cite{Homma1996}
\begin{equation}
S_{Ti}=1-\frac{V_{\mathbf{X}_{\sim i}}(E_{X_i}(Y|\mathbf{X}_{\sim i}))}{V(Y)},
\label{total}
\end{equation}
where $V_{\mathbf{X}_{\sim i}}(E_{X_i}(Y|\mathbf{X}_{\sim i}))$ measures the first order effect of $\mathbf{X}_{\sim i}$
on the model output which does not contain any effect corresponding to $X_i$.

In order to estimate the first order and total sensitivity indices,
a matrix combination approach is very common \cite{Saltelli_2008_Book}.
This approach calculates the conditional variance for each variable with a new sampling set.
In order to obtain a certain accuracy, this procedure requires often more than 1000 samples for each estimated conditional variance.
Thus, for models with a large number of variables and time consuming solver calls, this approach can not be applied efficiently.

\subsection{Coefficient of Correlation}
The coefficient of correlation is the standardized covariance between two random variables $X$ and $Y$
\begin{equation}
\rho(X,Y)=\frac{COV(X,Y)}{\sigma_X \sigma_Y},
\end{equation}
where $COV(X,Y)$ is the covariance and $\sigma$ is the standard deviation.
This quantity, known as the linear correlation coefficient, measures
the strength and the direction of a linear relationship
between two variables. It can be estimated from a given sampling set as follows
\begin{equation}
\rho(X,Y)\approx \frac{1}{N-1} \frac{\sum_{i=1}^N (x_i-\hat \mu_X)(y_i-\hat \mu_Y)}{\hat \sigma_X \hat \sigma_Y},
\end{equation}
where $N$ is the number of samples, $x_i$ and $y_i$ are the sample values, and $\hat \mu_X$ and $\hat \sigma_X$ 
are the estimates of the mean value
and the standard deviation, respectively.
The estimated correlation coefficient becomes more inaccurate, as its value is closer to zero,
which may cause a wrong deselection of apparently unimportant variables. 

If both variables have a strong positive correlation, the correlation coefficient is close to one.
For a strong negative correlation $\rho$ is close to minus one.
The squared correlation coefficient can be interpreted as the first order sensitivity index by assuming a linear dependence.
The drawback of the linear correlation coefficient is its assumption of linear dependence.
Based on the estimated coefficients only, it is not possible to decide on the validity of this assumption.
In many industrial applications a linear dependence is not the case. 
Correlation coefficients, which assume a higher order dependence or use rank transformations \cite{optislang2011}
solve this problem only partially.
Additionally, often interactions between the input variables are important.
These interactions can not be quantified with the linear and higher order correlation coefficients.

We can summarize that although the correlation coefficient can be simply estimated from a single sampling set,
it can only quantify first order effects with an assumed dependence without any quality control of this assumption.

\section{Polynomial based sensitivity analysis}
\subsection{Polynomial regression}
A commonly used approximation method is polynomial regression,
where the model response is generally approximated by a polynomial basis function
\begin{equation}
\mathbf{p}^T(\mathbf{x})=\left[ 1\;\; x_1\;\; x_2\;\; x_3\; \ldots \; x_1^2\;\; x_2^2\;\; x_3^2\; \ldots \; x_1x_2\;\; x_1x_3\; \ldots\; x_2x_3\; \ldots\right]
\end{equation}
of linear or quadratic order with or without coupling terms.
The model output $y_i$ for a given set $\mathbf{x}_i$ of the input parameters $\mathbf{X}$
can be formulated as the sum of the approximated value $\hat y_i$ and an error term $\epsilon_i$
\begin{equation}
y(\mathbf{x}_i) = \hat y_i(\mathbf{x}_i) + \epsilon_i = \mathbf{p}^T(\mathbf{x}_i)\boldsymbol{\beta} + \epsilon_i,
\end{equation}
where $\boldsymbol{\beta}$ is a vector containing the unknown regression coefficients.
These coefficients are generally estimated from a given set of
sampled support points by
assuming independent errors with equal variance at each point.
By using a matrix notation the resulting least squares solution reads
\begin{equation}
\boldsymbol{\hat \beta}=(\mathbf{P}^T\mathbf{P})^{-1}\mathbf{P}^T \mathbf{y},
\end{equation}
where $\mathbf{P}$ is a matrix containing the basis polynomials of the support point samples
and $\mathbf{y}$ is the vector of support point values.

\subsection{Coefficient of Determination}
\label{cod_section}
The Coefficient of Determination (CoD) can be used to assess the approximation quality of a
polynomial regression.
This measure is defined as the relative amount of variation explained by the approximation \cite{Montgomery2003}
\begin{equation}
R^2=\frac{SS_R}{SS_T}=1-\frac{SS_E}{SS_T}, \quad 0\leq R^2\leq 1,
\end{equation}
where $SS_T$ is equivalent to the total variation, $SS_R$ represents the variation due to the regression, and
$SS_E$ quantifies the unexplained variation,
\begin{equation}
SS_T=\sum_{i=1}^N(y_i-\mu_{Y})^2, \quad
SS_R=\sum_{i=1}^N(\hat y_i-\mu_{\hat Y})^2, \quad
SS_E=\sum_{i=1}^N(y_i-\hat y_i)^2.
\end{equation}
If the CoD is close to one, the polynomial approximation represents the support point values with small errors.
However, the polynomial model would fit exactly through the support points, if their number is equivalent to the number of coefficients $p$.
In this case, the CoD would be equal to one, independent of the true approximation quality.
In order to penalize this over-fitting, the adjusted Coefficient of Determination was introduced \cite{Montgomery2003}
\begin{equation}
R^2_{adj}=1-\frac{N-1}{N-p}(1-R^2).
\end{equation}
However, the over-estimation of the approximation quality can not be avoided completely.

\begin{figure}[th]
\begin{minipage}[b]{0.45\textwidth}
\includegraphics[width=\textwidth]{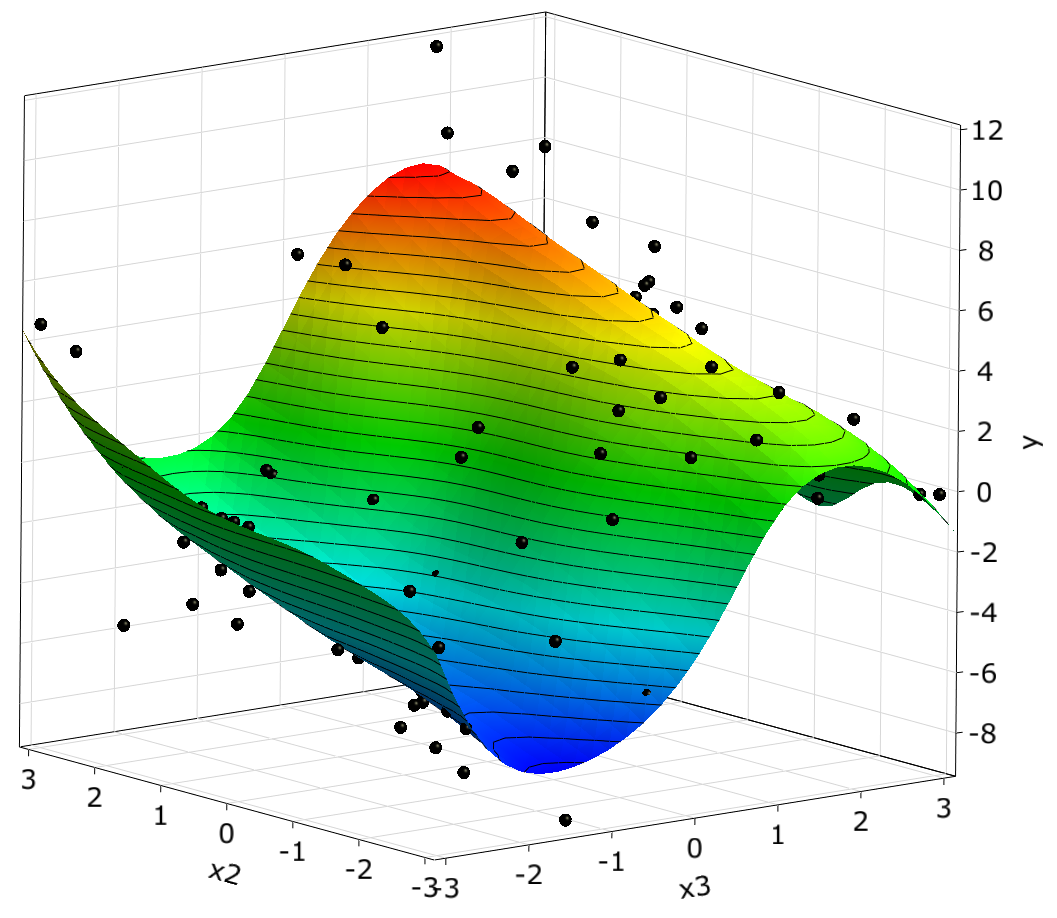}
\end{minipage}
\hfill
\begin{minipage}[b]{0.53\textwidth}
\includegraphics[width=1.1\textwidth]{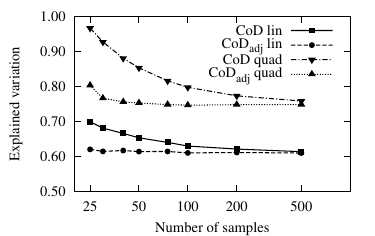}
\end{minipage}
 	\caption{Subspace plot of the investigated nonlinear function and convergence of the CoD measures with increasing number of support points}
	\label{COD}
\end{figure}
In order to demonstrate this  statement, an investigation of a nonlinear analytical function is performed.
The function reads in terms of five independent and uniformly distributed input variables as follows
\begin{equation}
Y=0.5X_1+X_2+0.5X_1 X_2 + 5.0 \sin(X_3) +0.2 X_4 + 0.1 X_5, \quad -\pi \leq X_i\leq \pi,
\label{coupled}
\end{equation}
where the contributions of the five inputs to the total variance are $X_1$: 18.0\%, $X_2$: 30.6\%, $X_3$: 64.3\%, $X_4$: 0.7\%, $X_5$: 0.2\%.
This means, that the three variables, $X_1$, $X_2$ and $X_3$, are the most important.

In Figure \ref{COD}, the convergence of the standard CoD of linear and quadratic response surfaces is shown, 
where a strong over-estimation of the approximation quality can be noticed, 
when the number of samples is relatively small. Even the adjusted CoD shows a similar behavior. 
This fact limits the CoD for cases where a large number of support points compared to the number of polynomial coefficients is available.
However, in industrial applications, we are in our interest, this is often not the case.
Another disadvantage of the CoD measure is its limitation to polynomial regression.
For other local approximation models, like interpolating Kriging, this measure may be equal or close to one,
however the approximation quality is still poor.

\subsection{Coefficient of Importance}
The Coefficient of Importance (CoI) was developed by Dynardo to quantify the input variable importance 
by using the CoD measure. Based on a polynomial model, including all investigated variables,
the CoI of a single variable $X_i$ with respect to the response $Y$ is defined as follows
\begin{equation}
CoI(X_i,Y) = CoI_{Y, X_i}=R^{2}_{Y,\mathbf{X}} - R^{2}_{Y,\mathbf{X}\sim i},
\end{equation}
where $R^{2}_{Y,\mathbf{X}}$ is the CoD of the full model including all terms of the variables in $\mathbf{X}$ and 
$R^{2}_{Y,\mathbf{X}\sim i}$ is the CoD of the reduced model, where all linear, quadratic and interactions terms belonging to $X_i$ are removed from
the polynomial basis. For both cases the same set of sampling points is used. If a variable has low importance,
its CoI is close to zero, since the full and the reduced polynomial regression model have a similar quality.
The CoI is equivalent to the explained variation with respect to a single input variable,
since the CoD quantifies the explained variation of the polynomial approximation.
Thus it is an estimate of the total effect sensitivity measure given in Equation \ref{total}.
If the polynomial model contains important interaction terms, the sum of the CoI values should be larger than the CoD of the full model.

Since it is based on the CoD, the CoI is also limited to polynomial models. If the total explained variation is over-estimated by the 
CoD, the CoI may also give a wrong estimate of the variance contribution of the single variables.
However, in contrast to the Coefficient of Correlation, the CoI can handle linear and quadratic dependencies including input variable interactions.
Furthermore, an assessment of the suitability of the polynomial basis is possible.
Nevertheless, an estimate of the CoI values using a full quadratic polynomial is often not possible because of the required large number of samples 
for high dimensional problems.

\section{Metamodel of Optimal Prognosis}
\subsection{Moving Least Squares approximation}
\begin{figure}[h!]
\begin{minipage}[c]{0.48\textwidth}
\includegraphics[width=1.05\textwidth]{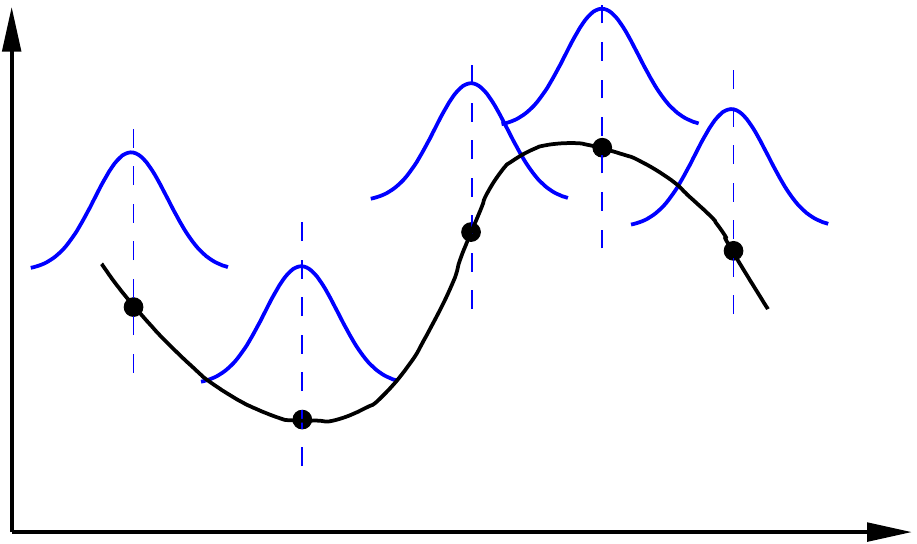}
\end{minipage}
\begin{minipage}[c]{0.50\textwidth}
\includegraphics[width=1.1\textwidth]{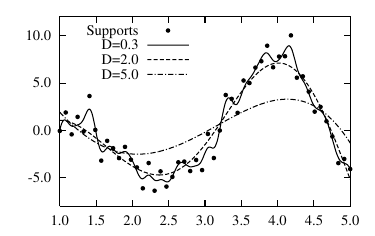}
\end{minipage}
 	\caption{Local weighting of support point values (left) and influence of the influence radius $D$ on the smoothing of the MLS approximation function (right)}
	\label{mls}
\end{figure}

In the Moving Least Squares (MLS) approximation \cite{Lancaster1981} a local character of the regression
is obtained by introducing 
position dependent radial weighting functions.
MLS approximation can be understood as an extension of the polynomial regression.
Similarly the basis function can contain every type of function, but generally only linear and quadratic terms are used.
The approximation function is defined as
\begin{equation}
\hat y(\mathbf x)=\mathbf{p}^T(\mathbf{x})\mathbf{a}(\mathbf{x}),
\end{equation}
with changing (``moving'') coefficients $\mathbf{a}(\mathbf{x})$ in contrast to the constant global coefficients of the polynomial regression.
The final approximation function reads
\begin{equation}
\hat y (\mathbf{x})=\mathbf{p}^T(\mathbf{x})(\mathbf{P}^T\mathbf{W(x)}\mathbf{P})^{-1}\mathbf{P}^T \mathbf{W(x)} \mathbf{y},
\end{equation}
where the diagonal matrix $\mathbf{W(x)}$ contains the weighting function values corresponding to each support point.
Distance depending weighting functions $w=w(\| \mathbf{x}-\mathbf{x}_i\|)$ have been introduced.
Mostly the well known Gaussian weighting function is used
\begin{equation}
w_{exp}(\| \mathbf{x}-\mathbf{x}_i\|) = exp\left(-\frac{\| \mathbf{x}-\mathbf{x}_i\|^2}{\alpha^2 D^2}\right),
\end{equation}
where the influence radius $D$ directly influences the approximation error.
A suitable choice of this quantity enables an efficient smoothing of noisy data. In Figure \ref{mls} the local weighting principle and the smoothing effect
is shown.

The MLS approach has the advantage that no training is necessary before an approximation point can be evaluated.
At each point only the weighting factors and the local polynomial coefficients have to be calculated.
This makes this method very fast compared to other approximation techniques.

\subsection{Coefficient of Prognosis}
In \cite{Most_2008_WOST} a model independent measure to assess the model quality was proposed.
This measure is the Coefficient of Prognosis (CoP), which is defined as follows
\begin{equation}
CoP=1-\frac{SS_E^{Prediction}}{SS_T},
\end{equation}
where $SS_E^{Prediction}$ is the sum of squared prediction errors. These errors
are estimated based on cross validation.
In the cross validation procedure, the set of support points is mapped to $q$ subsets. 
Then the approximation model is built by removing subset $i$ from the support points
and approximating the subset model output $\tilde y_j$ using the remaining point set.
This means that the model quality is estimated only at these points, which are not used to build the approximation model.
Since the prediction error is used instead of the fit,
this approach applies to regression and even interpolation models.

The evaluation of the cross validation subsets, which are usually between 5 and 10 sets, causes additional numerical effort 
in order to calculate the CoP. Nevertheless, for polynomial regression and Moving Least Squares, this additional effort
is still quite small since no complex training algorithm is required. For other meta-modeling approaches 
as neural networks, Kriging and even Support Vector Regression,
the time consuming training algorithm has to be performed for every subset combination. 

\begin{figure}[h!]
\centering
\includegraphics[width=0.6\textwidth]{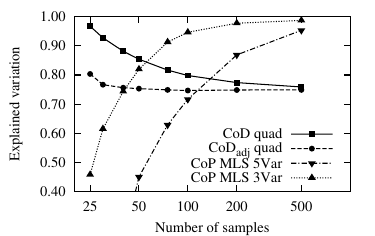}
 	\caption{Convergence of the CoP measure by using MLS approximation compared to the polynomial CoD measure}
	\label{COP}
\end{figure}

In Figure \ref{COP}, the convergence of the CoP values of an MLS approximation of the nonlinear coupled function 
given in Equation \ref{coupled} is shown in comparison to the polynomial CoD.
The figure indicates that the CoP values are not over-estimating the approximation quality as the CoD does for a small number of samples.
The influence radius of the MLS approximation is found by maximizing the CoP measure. 
As shown in Figure \ref{COP}, the convergence of the approximation quality is much better
if only the three important variables are used in the approximation model.

\subsection{Metamodel of Optimal Prognosis}
As shown in the previous section, the prediction quality of an approximation model may be improved if unimportant variables are removed
from the model.
This idea is adopted in the Metamodel of Optimal Prognosis (MOP) proposed in \cite{Most_2008_WOST}
which is based on the search for the optimal input variable set and
the most appropriate approximation model (polynomial or MLS with linear or quadratic basis).
Due to the model independence and objectivity of the CoP measure, it is well suited
to compare the different models in the different subspaces.
\begin{figure}[h!]
\centering
\includegraphics[width=0.6\textwidth]{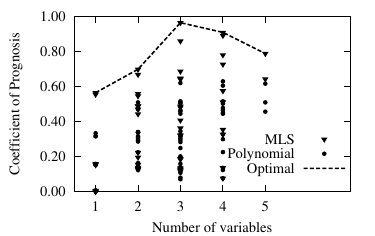}
 	\caption{CoP values of different input variable combinations and approximation methods obtained with the analytical nonlinear function}
	\label{MOP}
\end{figure}
In Figure \ref{MOP}, the CoP values of all possible subspaces and all possible approximation models
are shown for the analytical nonlinear function. The figure indicates that there exists an optimal compromise
between the available information, the support points and the model complexity, the number of input variables.
The MLS approximation by using only the three major important variables has a significantly higher CoP value than other combinations.
However for more complex applications with many input variables ,it is necessary to test a huge number of approximation models.
In order to decrease this effort, in \cite{Most_2008_WOST} advanced filter technologies are proposed,
which reduce the number of necessary model tests.
Nevertheless, a large number of inputs requires a very fast and
reliable construction of the approximation model. For this reason polynomials and MLS are preferred due to their fast evaluation.

As a result of the MOP, we obtain an approximation model, which includes the important variables.
Based on this meta-model, the total effect sensitivity indices, proposed in section \ref{saltelli},
are used to quantify the variable importance.
The variance contribution of a single input variable is quantified by the product of the CoP  and
the total effect sensitivity index estimated from the approximation model
\begin{equation}
CoP(X_i) = CoP \cdot S_{T}^{MOP}(X_i).
\end{equation}
Since interactions between the input variables can be represented by the MOP approach,
they are considered automatically in the sensitivity indices. If the sum of the single indices is significantly larger as the 
total CoP value, such interaction terms have significant importance.

\begin{figure}[h!]
\begin{minipage}[c]{0.49\textwidth}
\includegraphics[width=\textwidth]{figures/coupled_plot_x2_x3_mod}
\end{minipage}
\hfill
\begin{minipage}[c]{0.48\textwidth}
\includegraphics[width=\textwidth]{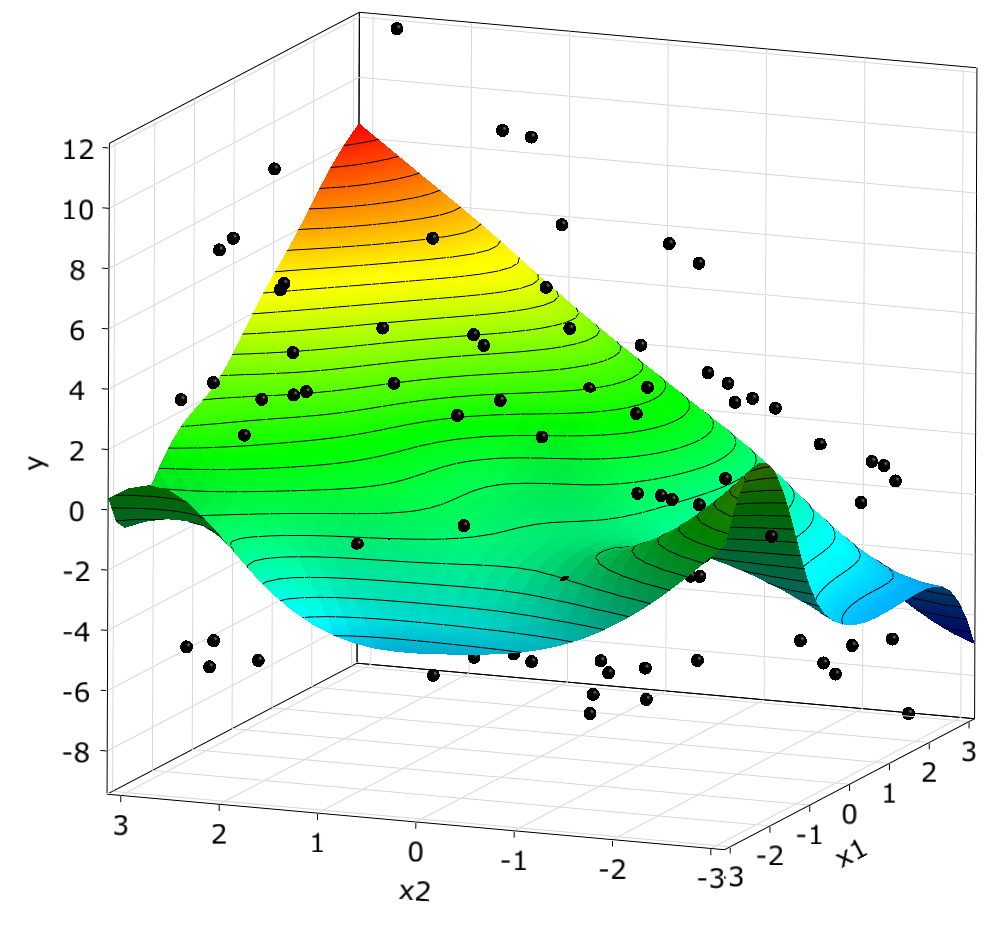}
\end{minipage}
 	\caption{$X_2$-$X_3$ and $X_1$-$X_2$ subspace plots of the MOP of the nonlinear analytical function}
	\label{MOP_plot}
\end{figure}

Additionally to the quantification of the variable importance, the MOP can be used to visualize the dependencies in 2D and 3D
subspaces. This helps the designer to understand and to verify the solver model.
In Figure \ref{MOP_plot} two subspace plots are shown for the MOP of the analytical test function. In the $X_2$-$X_3$ and $X_1$-$X_2$ subspace plots the
sinusoidal function behavior and the coupling term can be observed.
Additional parametric studies, such as global optimization can also be directly performed on the MOP.
Nevertheless, a single solver run should be used to verify the final result of such a parametric study or optimization.

\begin{figure}[h!]
\begin{minipage}[c]{0.48\textwidth}
\hspace{-5mm}
\includegraphics[width=1.05\textwidth]{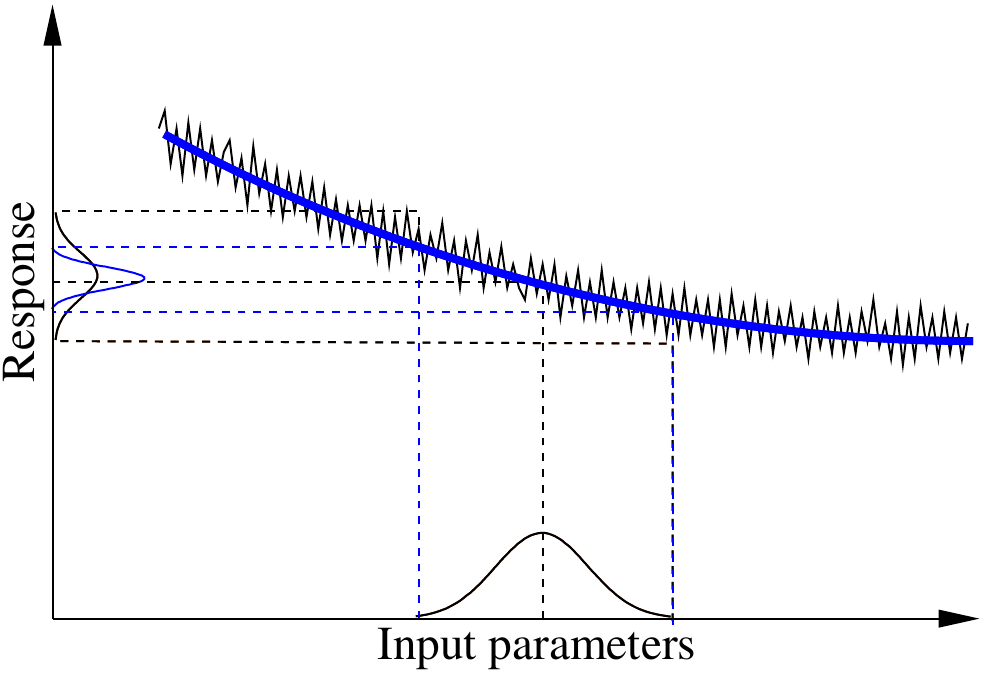}
\end{minipage}
\begin{minipage}[c]{0.50\textwidth}
\vspace{7mm}
\includegraphics[width=1.1\textwidth]{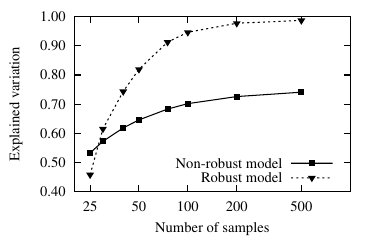}
\end{minipage}
 	\caption{Representation of a noisy model output by a smoothing approximation (left) and convergence behavior of the explained variation in the case of a
	robust and non-robust model (right)}
	\label{robustness}
\end{figure}

If the solver output contains unexplainable effects due to numerical accuracy problems,
the MOP approximation will smooth these noise effects as shown in Figure \ref{robustness}.
If this is the case, the CoP value of the MOP can be used to estimate the noise variation as the shortcoming 
of the CoP to 100\% explainability.
However, the unexplained variation may not be caused only by solver noise but also by a poor approximation quality.
This problem should be analyzed by increasing the number of samples in the case of low explainability.
If the CoP does not increase then it is an indicator for unexplainable solver behavior.

\section{Comparison with other approximation and selection methods}
\begin{figure}[h!]
\centering
\includegraphics[width=0.6\textwidth]{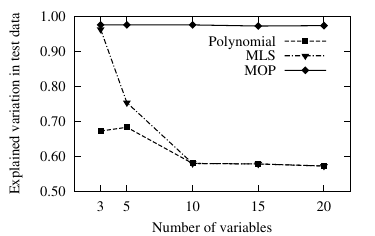}
 	\caption{Approximation quality for polynomial and MLS approximation compared to the MOP approach for the analytical function with increasing number of
	input variables}
	\label{compare_poly}
\end{figure}

In this section we compare the MOP approach with other approximation and 
variable selection methods.
Before using advanced meta-models, we investigate the test case with polynomials and Moving Least Squares.
The analytical nonlinear function introduced in section \ref{cod_section} is investigated 
by different numbers of input variables: only the three main important variables, all five variables and 
additional variables without any contribution.
In Figure \ref{compare_poly}, the explained variation of the polynomial and MLS approximation obtained with 100 support and 100 test points
is shown with respect to the total number of variables.
Due to the so-called curse of dimensionality the approximation quality decreases rapidly with increasing dimension.
If the MOP is applied, only the important variables are filtered out and the approximation is build in the optimal subspace.
This leads to a high approximation quality even for larger input dimensions.

In the next step, we investigate the Kriging approximation, that is also known as Gaussian process model,
which assumes a linear regression model similar to polynomial regression
\begin{equation}
y(\mathbf{x})= \mathbf{p}^T(\mathbf{x})\boldsymbol{\beta}+\epsilon(\mathbf{x}).
\end{equation}
Instead of independent errors, correlations between the error values are introduced by a spatial correlation function
similar to these in random fields
\begin{equation}
\mathbf{C}_{\boldsymbol{\epsilon\epsilon}}=\sigma^2 \boldsymbol{\Psi}, \quad
\Psi_{ij}=\exp(-\theta\|\mathbf{x}_i-\mathbf{x}_j\|^2),
\end{equation}
where $\mathbf{C}_{\boldsymbol{\epsilon\epsilon}}$ is the covariance matrix of the support points.
The exponential correlation function $\Psi$ uses often the quadratic norm of the spatial distance.

\begin{figure}[h!]
\centering
\includegraphics[width=0.6\textwidth]{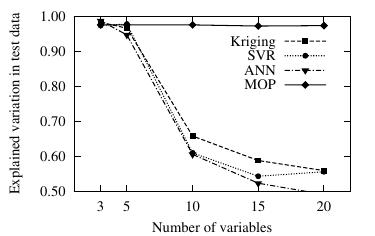}
 	\caption{Approximation quality for Kriging, Support Vector Regression (SVR) and Artificial Neural Networks (ANN) compared to the MOP approach for the analytical function}
	\label{compare_kernel}
\end{figure}

 A special case is called ordinary Kriging where only constant regression terms are used
\begin{equation}
y(\mathbf{x})= \mu+\epsilon(\mathbf{x}).
\end{equation}
For this case the approximation function reads
\begin{equation}
\hat y(\mathbf{x}) = \hat \mu + \boldsymbol{\psi}(\mathbf{x})^T\boldsymbol{\Psi}^{-1}(\mathbf{y}-\mathbf{1}\hat \mu)=
\hat \mu + \boldsymbol{\psi}(\mathbf{x})^T\mathbf{w}
\end{equation}
The optimal correlation parameters are obtained generally by the maximum likelihood approach or by cross validation.
In our study we use the additional test data set, however, the cross validation approach is more robust.
The determination of the Kriging weights $\mathbf{w}$ requires the inversion of the Gram matrix $\boldsymbol{\Psi}$,
which is very time consuming for a large number of support points.
For this reason, the cross validation procedure requires a significantly higher numerical effort as for Kriging when compared to the MLS approach.
In Figure \ref{compare_kernel}, the approximation quality of the ordinary Kriging approach is shown for the test function.
The significant decrease of the explained variation is similar to that of the MLS approach.

Furthermore, Support Vector Regression (SVR) and Artificial Neural Networks (ANN) are investigated. A detailed presentation of these methods can be found in
\cite{Roos_2007_WOST}. The obtained results are shown additionally in Figure \ref{compare_kernel}, which show a similar behavior as in Kriging and MLS.
All the presented results show that the utilization of complex meta-model approaches will not overcome the curse of dimensionality.
However, the MOP enables  the determination of the optimal variable subspace by using fast and reliable approximation methods. In 
the most cases this variable reduction leads to a significantly better approximation quality.

\begin{figure}[h!]
\begin{minipage}[b]{0.48\textwidth}
\hspace{-5mm}\includegraphics[width=1.15\textwidth]{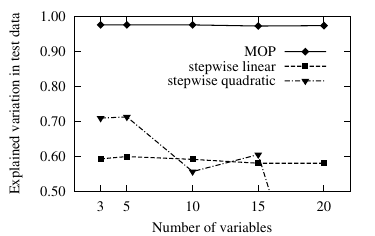}
\end{minipage}
\hfill
\begin{minipage}[b]{0.48\textwidth}
\hspace{-3mm}\includegraphics[width=1.15\textwidth]{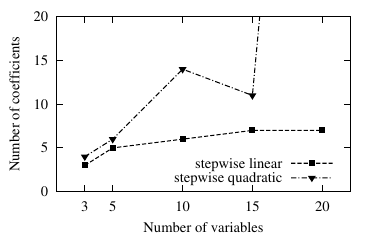}
\end{minipage}
 	\caption{Approximation quality and variable selection of MATLAB's stepwise regression approach compared to the MOP results}
	\label{matlab}
\end{figure}

Finally the MOP approach is compared to the polynomial stepwise selection method. In this approach polynomial coefficients are selected by different importance criteria
in order to detect the important variables. For the comparison, the state of the art implementation in \cite{Matlab2010}
is used. In this implementation important polynomial coefficients are selected by F-test statistics based on a given polynomial degree. The results given in 
Figure \ref{matlab} 
indicate that the selection procedure works appropriately only for a linear polynomial basis. By using a full quadratic basis, the number of 
selected coefficients increases dramatically and the approximation quality decreases.
This example clarifies the power of the prediction based variable selection applied inside the MOP approach.

\section{Application in Noise Vibration Harshness analysis}
\begin{figure*}
\center
	\includegraphics[width=0.7\textwidth]{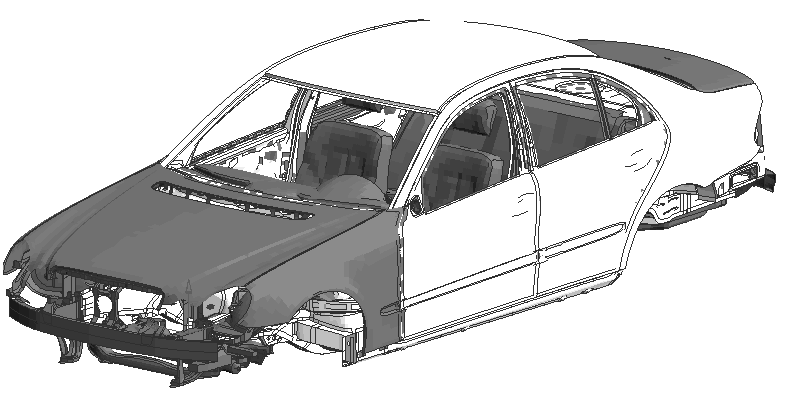}
	\includegraphics[width=0.7\textwidth]{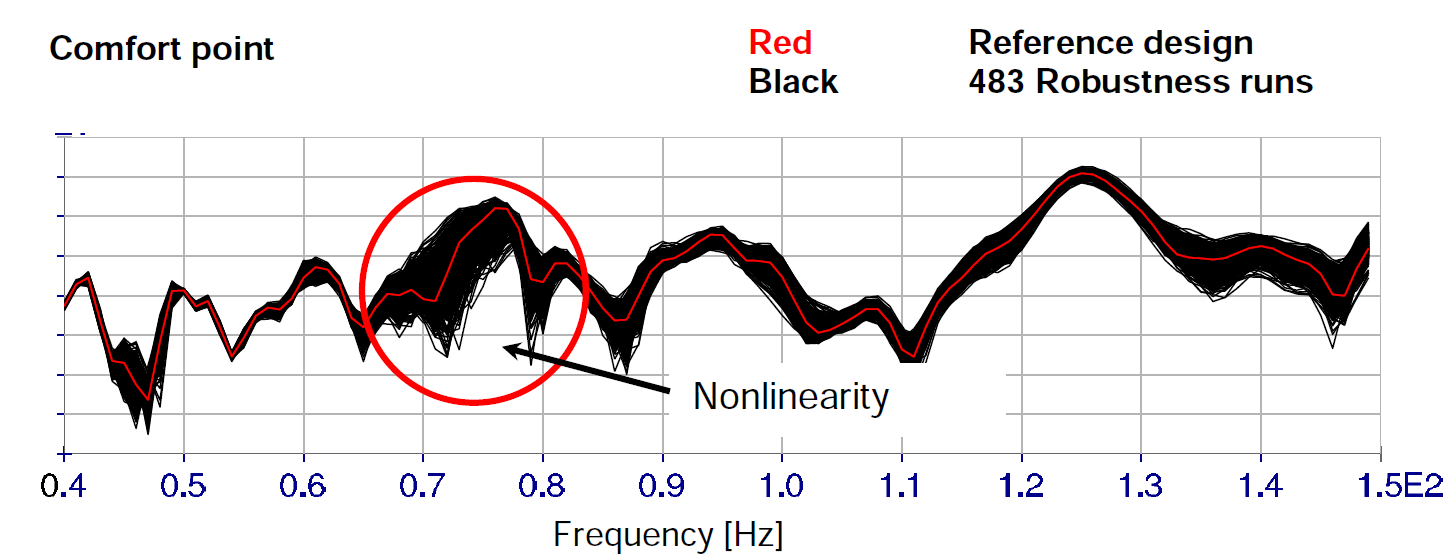}
\caption{Car body with varied 46 sheets thicknesses and investigated sound pressure level depending on the frequency}
\label{car_body}
\end{figure*}

In this example we apply the MOP approach in the framework of a robustness analysis
in automotive industry.
Here we investigate an example presented in \cite{Will_2004_VDI}
where the Noise Vibration Harshness (NVH) is analyzed.
Input parameters in this analysis are 46 sheet thicknesses of a car body
which are varied within a +/- 20\%.
The sound pressure levels at certain frequencies are the outputs which are investigated.
In Figure \ref{car_body}, the car body including the sheets with varying thicknesses are shown.

\begin{figure*}
\center
	\includegraphics[width=0.49\textwidth]{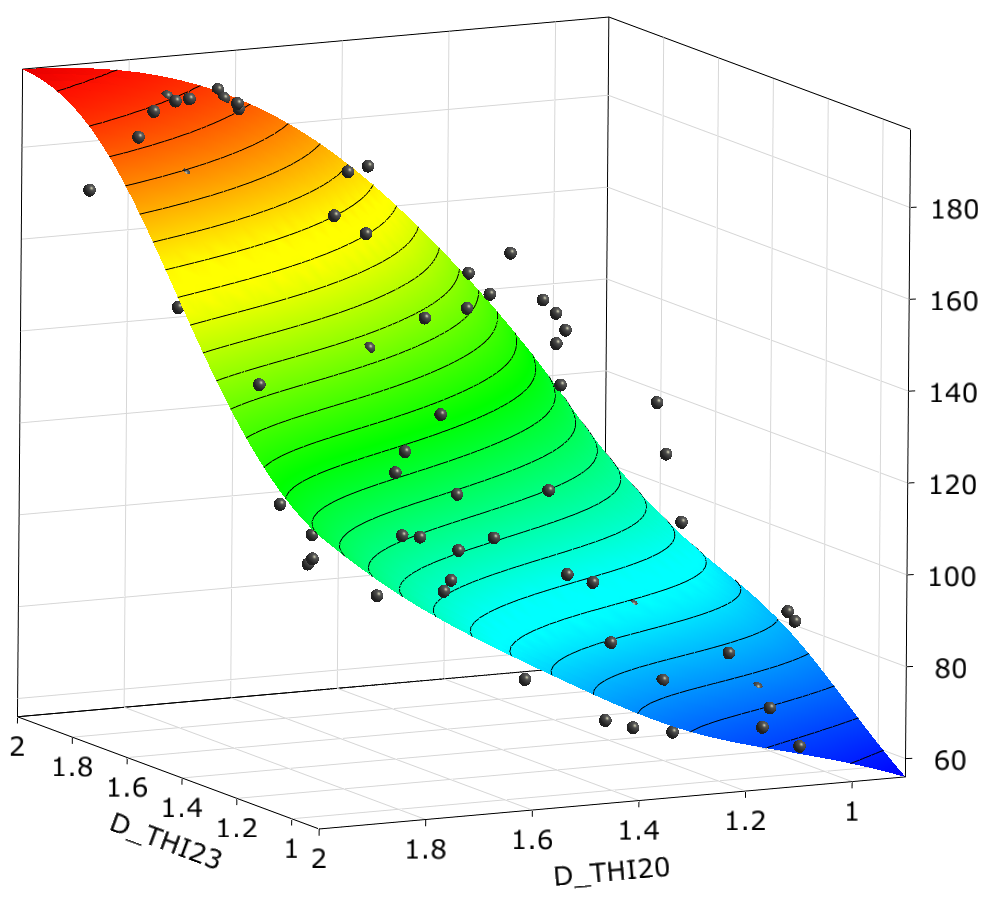}
	\includegraphics[width=0.49\textwidth]{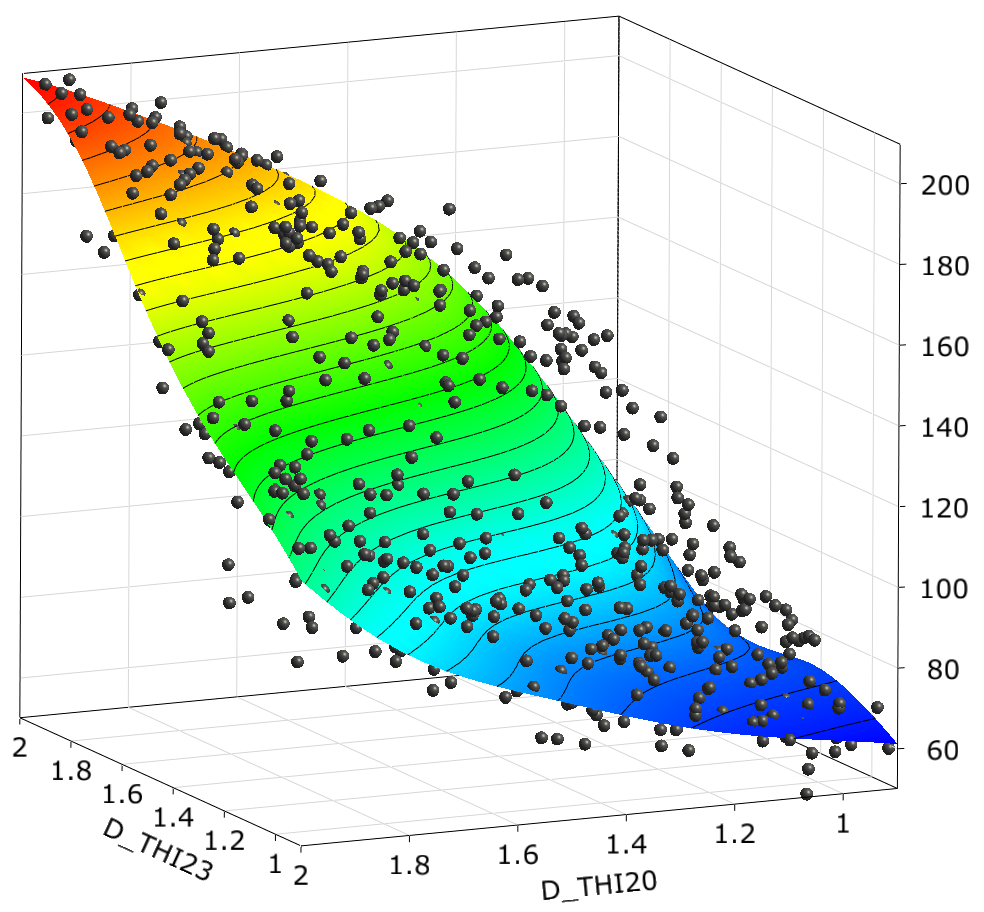}
\caption{Approximation function in the subspace of the two most important inputs using 100 samples as supports (left) 
and 800 samples (right)}
\label{mop_functions}
\end{figure*}

In order to find the important parameters with a small number of samples, which are obtained from very time consuming finite element simulations,
the application of MOP is very attractive.
Based on the MOP, the total effect sensitivity indices are calculated.
In table \ref{sensitivity_DC}, the resulting indices of one sound pressure value including 
the approximation quality 
are given for different numbers of samples obtained from Latin Hypercube Sampling. 
\begin{table}[h]
\footnotesize
\begin{center}
\begin{tabular}{lccccccc}
   \hline
No. samples		&100	&200	 &400	  &600     &800\rule[0mm]{0pt}{2.5ex}\\
   \hline
$CoP$	&90.9\%	&91.7\%   &95.7\%   &96.3\%   &96.9\%\rule[0mm]{0pt}{2.5ex}\\
$CoP(X_5)$		&-		&-  	 &2.4\%   &2.3\%   &2.7\%\rule[0mm]{0pt}{2.5ex}\\
$CoP(X_6)$		& 6.0\%  &5.3 \%  &8.2\%   &8.3\%   &8.7\%	\\
$CoP(X_{20})$	&41.3\%  &42.7\%   &42.3\%   &43.4\%   &42.2\%\\     
$CoP(X_{23})$	&49.1\%  &48.0\%   &50.7\%   &51.0\%   &53.8\%\\  		   
   \hline
\end{tabular}
\end{center}
\caption{Convergence of approximation quality and total sensitivity indices for the most important sheet thickness}
\label{sensitivity_DC}
\end{table}
The table indicates that even for a very small number of samples compared to the number of input variables 
the most important variables can be detected. When increasing the number of samples, 
additional minor important inputs are detected and considered in the optimal meta-model.
The influence of coupling terms increases from approximately 5\% to 11\% 
due to the better approximation quality. In figure \ref{mop_functions}
the approximation functions are shown in the subspace of the two most important variables $X_{20}$ and $X_{23}$.
The figure indicates that with only 100 samples the general functional behavior can be represented.

Due to the efficiency of the proposed sensitivity analysis even for nonlinear coherences between inputs and outputs,
the MOP approach is applied in several industrial projects in cooperation with the German automotive industry. 

\section{Summary}
In this paper the Metamodel of Optimal Prognosis was presented. It was shown, that in contrast to one-dimensional
correlation coefficients and multi-dimensional polynomial based sensitivity analysis, the MOP enables an efficient and reliable
estimation of the input variable importance. With the help of the objective quality measure CoP, the MOP approach detects
the subset of most important variables necessary for the optimal approximation. 
The application of variable selection proves to be successful even for higher dimensional problems where the full space 
approximations using advanced meta-models show weak performance.

The determined optimal approximation function of one or more model responses can be used further to get a first estimate of a global optimum.
In the framework of robustness analysis, the CoP value of the MOP is a useful estimate of possible solver noise and an indicator for the
usability of the investigated CAE model.

\newpage


\end{document}